\documentclass[twocolumn]{openjournal}

\usepackage{xcolor}
\usepackage{textgreek}
\usepackage[utf8]{inputenc}
\usepackage[english]{babel}

\usepackage{hyperref}
\hypersetup{
    unicode, 
    colorlinks=true,
    linkcolor=linkcolor,
    citecolor=linkcolor,
    filecolor=linkcolor,
    urlcolor=linkcolor,
}
\usepackage{color,colortbl}
\definecolor{linkcolor}{rgb}{0.0,0.3,0.5}
\usepackage{tensind}
\tensordelimiter{?}
\DeclareGraphicsExtensions{.bmp,.png,.jpg,.pdf}
\usepackage{verbatim}
\usepackage[normalem]{ulem}
\usepackage{orcidlink}
\usepackage{soul}
\usepackage{graphicx}
\usepackage{multirow}
\usepackage{amsmath}

\DeclareRobustCommand{\VAN}[3]{#2}
\let\VANthebibliography\thebibliography
\def\thebibliography{\DeclareRobustCommand{\VAN}[3]{##3}\VANthebibliography}

\graphicspath{ {./figs/} }

\newcommand{\gs}{\mathrel{\lower0.6ex\hbox{$\buildrel {\textstyle >}
 \over {\scriptstyle \sim}$}}}
\newcommand{\ls}{\mathrel{\lower0.6ex\hbox{$\buildrel {\textstyle <}
 \over {\scriptstyle \sim}$}}}

\newcommand{\lta}{\mathrel{\spose{\lower 3pt\hbox{$\mathchar"218$}}
     \raise 2.0pt\hbox{$\mathchar"13C$}}}
\newcommand{\gta}{\mathrel{\spose{\lower 3pt\hbox{$\mathchar"218$}}
     \raise 2.0pt\hbox{$\mathchar"13E$}}}

\newcommand{\oiiia}{\mbox{[O\,{\textsc{iii}}]}\,}

\newcommand{\halpha}{\mbox{H\,{\sc$\alpha$}}\,}
\newcommand{\hbeta}{\mbox{H\,{\sc$\beta$}}}

\begin{document}
\title{Making the most of pure parallels: Machine learning augmented photometric redshifts for sparse \emph{JWST} filter sets}

\author{Kenneth J. Duncan \orcidlink{0000-0001-6889-8388}}
\email{kdun@roe.ac.uk}
\affiliation{Institute for Astronomy, University of Edinburgh, Royal Observatory, Blackford Hill, Edinburgh, EH9 3HJ, UK}


\begin{abstract}
Photometric redshifts (photo-$z$s) are an essential tool for galaxy evolution science with \emph{JWST}.
However, for deep surveys with more limited filter sets (i.e. $N_{\text{filt}} \sim6$) such as large pure parallel surveys, the most commonly used template-fitting based photo-$z$ approaches can yield highly confident but spurious results for high-$z$ populations of interest.
The utility and legacy value of these datasets could therefore be negatively impacted.
To address this challenge, we present an application of machine learning (ML) based photo-$z$ techniques to deep \emph{JWST} photometric datasets.
We employ two different ML algorithms, using Gaussian processes and nearest-neighbour estimates, alongside a more standard template fitting approach.
We show that simple nearest-neighbour based estimates can provide more accurate photo-$z$s than template fitting out to $z\sim8$, as well as reducing the fraction of catastrophic outliers by a factor of $\sim2-3$.
Additionally, `hybrid' estimates combining template and ML can yield further improvements in overall accuracy and reliability while retaining some ability to predict photo-$z$ out to $z > 10$. 
The nearest-neighbour only or hybrid estimates can achieve photo-$z$s with robust scatter of $\sigma_{\text{NMAD}}\sim0.03-0.04$ and outlier fractions of $\sim3-10\%$ between $0 < z \lesssim 8$ from just 6 NIRCam bands, with negligible additional computational costs compared to standard template fitting.
Our methodology is easily adaptable to alternative datasets, filter combinations or training samples.
Overall, our results highlight the potential for even simple ML techniques to enhance the scientific return of \emph{JWST} pure parallel and wide-area surveys.
\end{abstract}


\maketitle

\section{Introduction}
\label{sec:intro}
With the launch of \emph{JWST} and the unprecedented spectroscopic sensitivity and wavelength coverage of the Near-Infrared Spectrograph \citep[NIRSpec;][]{Jakobsen2022}, it is now possible to obtain robust spectroscopic redshifts for galaxies at almost any redshift \citep[e.g. $z > 12$][]{curtis-lake2023,arrabalharo2023,Carniani2024} or apparent magnitude probed by corresponding photometric surveys \citep{Eisenstein2023, bunker2023}.
Nevertheless, photometric redshift estimates (photo-$z$s) still remain a critical tool in studies of galaxy formation and evolution, both for initial identification and selection of the relevant spectroscopic targets of interest \citep[e.g.][]{hu2024,Maseda2024}, and for enabling statistical studies of complete samples \citep[e.g.][]{Adams2024, Donnan2023,Donnan2024,Finkelstein2024,McLeod2024}.

For galaxy evolution studies at moderate to high-redshifts (i.e. $z \gtrsim 1$), where samples are drawn from deep survey fields with a large number of observed photometric filters, template fitting approaches that fit model or empirical template libraries to the observed photometry remain the most effective and popular approach for estimating photo-$z$s \citep[e.g.][]{Brammer2008}.
Despite some high profile surprises \citep[c.f][]{Donnan2023, arrabalharo2023}, the first few years of \emph{JWST} galaxy surveys has seen wide-ranging success in the application of template photo-$z$s to these new regimes.
To achieve this, active effort has been invested in optimising the templates used for \emph{JWST} photo-$z$ studies; through the inclusion of younger stellar populations more appropriate for the early Universe \citep{larson2023b} and incorporating new populations of active galactic nuclei \citep[AGN, e.g.][]{Killi2023}.

For the best studied extra-galactic legacy fields, where \emph{JWST} observations spanning the full Near-infrared Camera \citep[NIRCam;][]{Rieke2005, Rieke2023} wavelength coverage are building upon extensive UV and optical photometry from \emph{Hubble Space Telescope} (HST), photo-$z$ estimates can reach near spectroscopic precision \citep{naidu2024,duncan2025}.
However, thanks to pure parallel surveys that are designed to maximise \emph{JWST}'s observing efficiency \citep[][]{williams2024, morishita2025}, and wide-area surveys in novel survey fields \citep{NEXUS2024}, an increasingly large fraction of \emph{JWST}/NIRCam imaging now exists in survey fields without comparably deep optical observations.
Even with the high signal-to-noise provided by \emph{JWST} and the latest template libraries, the more limited and sparse wavelength coverage available can lead to significant photo-$z$ degeneracies \citep{williams2024}, potentially hampering the scientific return from these valuable datasets and mitigating the benefits offered by the increased survey volumes and reduced cosmic variance.

Here, machine learning (ML) approaches to photo-$z$ estimation could provide a possible solution.
ML photo-$z$ estimates that use regression or classification algorithms trained on reference spectroscopic redshift samples \citep[e.g.][]{Collister2004fx, CarrascoKind:2013kd, CarrascoKind:2014gb} have become the preferred approach for wide area photometric surveys \citep[see e.g.][]{2021MNRAS.501.3309Z, duncan2022}, especially in the context of next generation cosmology surveys \citep{2020A&A...644A..31E,newman2022}.
The difficulty of obtaining \emph{representative} training samples (spanning the necessary redshift, magnitude and colour space) for deep survey fields means that such approaches have not traditionally been competitive. 
However, the extraordinary wealth of both literature spectroscopic surveys in the legacy fields, along with the growing numbers of \emph{JWST}/NIRSpec multi-object and slit-less spectroscopic samples extending to fainter magnitudes and higher redshifts, mean that these fields could now potentially provide the necessary training data to enable competitive ML estimates for \emph{JWST} survey fields with more limited filters or no comparably deep optical observations.

Crucially, it is important to note that these ML photo-$z$ estimates do not have to entirely replace or supersede template-based estimates in these fields.
\citet{Duncan2019,Duncan2019b} and \citet[][see also \citeauthor{2017MNRAS.466.2039C}~\citeyear{2017MNRAS.466.2039C}]{Duncan:21} previously demonstrated on multiple observational datasets that `hybrid' photo-$z$ estimates that combine estimates from both ML and template fitting approaches can yield better overall accuracy and precision than either input methodology alone.

In this paper, we explore the potential for ML photo-$z$ estimates to supplement template estimates in wide-area \emph{JWST} surveys with uniform but relatively sparse filter coverage.
We explore two ML-based approaches to photo-$z$ estimation for uniformly processed \emph{JWST}/NIRCam photometric catalogues that maximise the spectroscopic samples available for training.
Firstly, building on the methodology successfully applied by \citet{duncan2022} to all-sky photometric survey data, we apply the Gaussian Process redshift code \textsc{GPz} \citep{2016MNRAS.455.2387A}.
Secondly, we also explore a simple nearest-neighbour based photo-$z$ technique motivated by the approach employed for \emph{Euclid} \citep{2020A&A...644A..31E, euclid_q1_photoz}. 
Finally, motivated by the demonstrated success of combining template and \textsc{GPz} estimates in `hybrid' consensus photo-$z$s for specific galaxy populations \citep{Duncan:2017ul, Duncan2019, Duncan:21}, we then explore the results achieved from the combined predictions of ML and template fitting based photo-$z$ estimates.

Given that spectroscopic survey datasets from \emph{JWST} are rapidly evolving (and hence the available training samples), the `best' results we present here for a given set of photometric filters will be quickly superseded.
Similarly, we cannot exhaustively present models for all potential filter combinations.
We therefore provide the full code for our analysis, including a dedicated \textsc{Python} package for simplifying the application of \textsc{GPz}, and annotated notebooks presenting the exact analysis outlined in this paper.
The methodology is easily reproducible, and extendable to different photometric datasets (i.e. filter combinations) or new spectroscopic training samples.

The remainder of the paper is set out as follows.
In Section~\ref{sec:data} we outline the \emph{JWST}/NIRCam photometry catalogues used for training and prediction, as well as the details of the spectroscopic training and test samples.
Section~\ref{sec:method} then presents the \textsc{GPz} photo-$z$ methodology employed, highlighting differences with respect to previous applications \citep{duncan2022}.
Next, in Section~\ref{sec:results} we present and analyse the resulting photo-$z$ estimates for both individual methods and for consensus estimates combining both ML and template fitting approaches.
In Section~\ref{sec:discussion} we discuss the broader implications and potential for the results presented.
Finally, in Section~\ref{sec:concl} we summarise our results and conclusions.
Throughout this paper, all magnitudes are quoted in the AB system \citep{1983ApJ...266..713O} with a photometric zero-point $f_{0} = 3631~\rm{Jy}$, unless otherwise stated. We also assume a $\Lambda$ Cold Dark Matter cosmology with $H_{0} = 70$ km\,s$^{-1}$\,Mpc$^{-1}$, $\Omega_{m}=0.3$ and $\Omega_{\Lambda}=0.7$.

\section{Data}\label{sec:data}
\subsection{JWST/NIRCam Photometric Catalogues}
The approach presented in this paper is intended to be agnostic to the details of the photometric catalogues used and the results are not strongly dependent on the exact choice of photometry, provided that they are processed in a consistent and uniform way.
For this analysis we therefore use a subset of the uniformly processed photometry catalogues available through the DAWN \emph{JWST} Archive (DJA)\footnote{\url{https://dawn-cph.github.io/dja/index.html}}, which includes the majority of available NIRCam data in the best studied extragalactic fields 

As the fields with the largest number of available literature spectroscopic samples, and also the targets of large NIRSpec spectroscopic follow-up, we include the DJA catalogues for the following fields:
\begin{itemize}
    \item The Abell 2744 field, including GLASS \citep{treu2022} and UNCOVER \citep{bezanson2024} data. We include both the cluster and parallel fields (`abell2744clu' and `abell2744par' respectively in the DJA), with the assumption that strong lensing impacts on the apparent magnitude are negligible w.r.t. luminosity distance for the majority of sources.
    \item The Cosmic Evolution Early Release Science \citep[CEERS;][]{finkelstein2023} field, located within the Extended Groth Strip (EGS).
    \item \emph{JWST} Advanced Deep Extragalactic Survey \citep[JADES;][]{Eisenstein2023} imaging in both the GOODS North and South fields.
    \item Deeper imaging in the GOODS-South region from the Next Generation Deep Extragalactic Exploratory Public \citep[NGDEEP;][]{bagley2024} survey. 
    \item Both UDS and COSMOS fields from the `Public Release Imaging For Extragalactic Research' (PRIMER; Dunlop et al. \emph{in prep}).
\end{itemize}

The DJA reductions of the NIRCam imaging use the \texttt{Grizli} \citep{grizli} package, with additional details outlined by \citet{valentino2023}.
In summary, the pipeline includes corrections for cosmic rays and stray light, as well as updated sky flats compared to the standard pipeline processing.
All imaging is aligned to the Gaia DR3 astrometry scale \citep{Gaia2022} and drizzled onto a common pixel scale of 0.02\arcsec for the short wavelength (SW) NIRCam channels and 0.04\arcsec for the long wavelength (LW) channels.
Source detection is performed on stacks of the available LW channels, typically F277W+F356W+F444W.

From the DJA photometric catalogues we use the 0.5\arcsec\, diameter aperture photometry in each band, corrected to total flux based on the ratio between the aperture flux and an elliptical Kron aperture \citep{kron1980} on the LW detection image.
The measured flux densities are then corrected for the negligible impacts of galactic extinction.
We extract the galactic extinction, $E(B-V)$, for each source from the \citet{sfd1998} maps, queried through the \texttt{dustmaps} \citep{2015ApJ...810...25G} package, with the corresponding extinction in each filter, $A_{\lambda}/E(B-V)$, derived from the \citet{1999PASP..111...63F} Milky Way extinction curve.
Finally, when generating a combined photometric catalogue for a given set of observed NIRCam filters, we only include sources with observations in all of the relevant filters, free from significant photometric flagging and with physical half-light radii (i.e. excluding residual cosmic rays/artefacts).

For the primary analysis presented below, all analysis is performed for a sample limited to only the six most common NIRCam filters available in the wide area PANORAMIC \citep{williams2024} and BEACON \citep{morishita2025} pure parallel surveys, specifically the F115W, F150W and F200W SW channels, as well as the F277W, F356W and F444W LW channels. 

\subsubsection{\emph{asinh} magnitudes}
The \emph{JWST} photometric catalogues outlined above provide forced flux density measurements in all available filters for a given source.
For many sources this leads to non-detections in one or more filters, with some of those non-detections being adding significant information to the potential photo-$z$ solution.
The colour information from non-detections is particularly important for high redshift studies where absence of flux below the Lyman break feature provides one of the key observational signatures for high-$z$ galaxies.
Although we note that for datasets consisting purely of JWST/NIRCam observations, the Lyman break is only relevant at $z\gtrsim7$ (when F090W is available) or $z\gtrsim8.5$ (if F115W and redder).

Following \citet{duncan2022} and other works, we therefore also derive \emph{asinh} magnitudes \citep[or `luptitudes';][]{1999AJ....118.1406L} for use as input features for training and prediction of photo-$z$ estimates with \textsc{GPz}.
For a given flux density, $f_{\nu}$, with photometric zeropoint $f_{0} = 3631~\rm{Jy}$), the \emph{asinh} magnitudes are defined as
\begin{equation}
	 m = \frac{-2.5}{\log(10)} \times \sinh^{-1}\left ( \frac{f_{\nu}/f_{0}}{2b} \right ) + \log(b)
\end{equation}
where, $b$, known as the softening parameter, is defined relative to $f_{0}$. 
Corresponding magnitude uncertainties are then given by 
\begin{equation}
    \sigma_{m} =  \frac{-2.5}{\log(10)} \times \frac{(\sigma_{f_{\nu}}/|f|)}{ \sqrt{\left(1 + (2b / (f_{\nu}/f_{0}))^{2}\right )}}
\end{equation}
where $\sigma_{f_{\nu}}$ is the uncertainty on the measured flux density (and $b$/$f_{0}$ as above).

As demonstrated previously by \citet{2019MNRAS.489..820B} and \citet{duncan2022}, photo-$z$ estimates derived using \emph{asinh} magnitudes are not strongly sensitive to the exact softening parameter ($b$) assumed when the depth of the data is relatively homogeneous within a field.
However, given the large potential variation in depth between \emph{JWST} NIRCam survey fields, we cannot assume homogeneity across the full dataset for a given filter.
We therefore derive the softening parameter on a per-field basis for each filter, based on the 1$\sigma$ background noise from apertures across the field.
In future studies, if systematic effects from this assumption become the dominant limitation on the accuracy of predicted photo-$z$s, a more detailed approach that calculates softening parameters based on local noise within the field could be employed.


\subsection{Spectroscopic Training Sample}
We construct a parent spectroscopic training sample from three primary sources of redshifts.
Firstly, given the designed overlap between our selected \emph{JWST} observations and previous legacy surveys, there exists a substantial array of literature spectroscopic follow-up in these fields from decades of both small and large-scale observing campaigns.
We therefore start from the compilation of spec-$z$s in the five Cosmic Assembly Near-Infrared Deep Extragalactic Legacy Survey \citep[CANDELS;][]{Grogin2011,Koekemoer:2011br} fields presented by \citet{Kodra2023}.
We include all available spec-$z$s at cosmological redshifts ($z > 0.01$), and include additional grism redshifts where available.
In total, this yields 12\,626 unique sources across the five CANDELS fields (COSMOS, EGS, GOODS-N/S, UDS).\footnote{We note that larger literature samples exist in these fields, with CANDELS compilations potentially adding a further $\sim5000$ additional training sources \citep[c.f][]{Duncan2019b}. These additional sources do not significantly improve the results, we therefore opt for the public compilation to ensure the analysis can be fully reproduced.}

Our second major source of training spec-$z$s comprises publicly available JWST/NIRSpec spectroscopic confirmations also available through the DJA.\footnote{\url{https://s3.amazonaws.com/msaexp-nirspec/extractions/nirspec_public_v4.4.html}}
Similar to the photometric catalogues, the DJA spectroscopic sample consists of observations from a broad range of survey programmes uniformly processed with the \texttt{msaexp} package \citep[][with additional details presented in \citeauthor{heintz2024}~\citeyear{heintz2024}]{msaexp}. 
We include only those spec-$z$s graded as robust ($\texttt{grade}=3$) and remove duplicate entries where multiple NIRSpec observations of the same object have been taken by keeping the highest median signal-to-noise ratio (SNR) sources.
We note that higher resolution observations may yield more precise spec-$z$s despite lower SNR, however for the purposes of photo-$z$ training we expect the spec-$z$ precision to be sufficient (and the potential for catastrophic failures reduced due to the larger wavelength coverage).

After these cuts, the resulting sample of unique NIRSpec spec-$z$s across the full sky totals 20\,311 sources.
The full DJA compilation consists of spectra and associated spectroscopic redshifts from many individual programmes, however the bulk of the sample comes from a number of large programmes presented in \citet{finkelstein2023}, \citet{carnall2024},  \citet{DEugenio2025}, \citet{Maseda2024}, \citet{bezanson2024}, \citet{degraaff2024} and \citet{nanayakkara2025}.

Finally, we also make use of \emph{JWST} derived spec-$z$s from slit-less spectroscopic surveys, including additional sources ($\texttt{q}=3$) from the \halpha \citep[][144 unique sources]{covelopaz2024} and \oiiia+\hbeta\, samples \citep[][28]{Meyer2024} from the FRESCO survey in GOODS North and South \citep{Oesch2023}, as well as robust line emitters from the `All the Little Things' \citep[ALT;][1406]{naidu2024} survey over the Abell 2744 cluster.

When combining the samples together, we check for duplicates within a 0.4\arcsec\, matching radius.
Where available, we assume \emph{JWST}/NIRSpec redshifts are the most robust due to the availability of unambiguous rest-frame optical/near-infrared features at all redshifts, followed by \emph{JWST}/WFSS and then the CANDELS literature compilation.
This results in a parent training sample of 33\,608 unique spec-$z$ sources.

However, since the underlying CANDELS fields typically cover larger areas than their associated NIRCam observations, and the orientation of \emph{JWST} parallels making matching NIRCam and NIRSpec survey footprints extremely difficult, we expect many of these sources will fall outside the coverage for our photometric sample (or with partial filter coverage).
Matching to our fiducial photometric catalogue using the same 0.4\arcsec\, maximum separation yields a total of 19\,056 training sources.

\begin{figure}[t!]
    \centering
    \vspace{0cm}
    \includegraphics[width=\columnwidth]{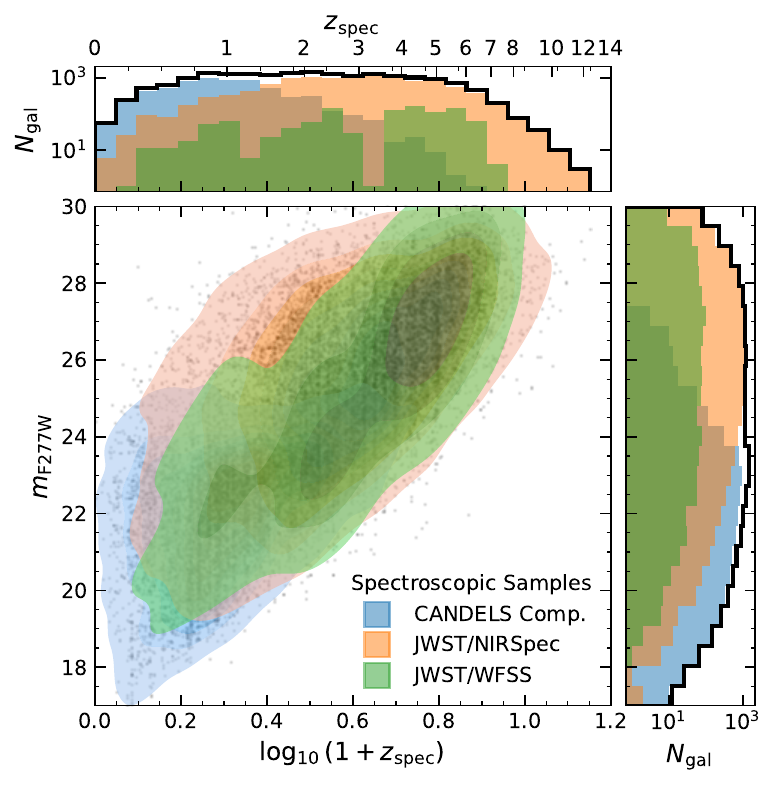}
    \caption{Redshift and F277W magnitude ($m_{\text{F277W}}$) distribution for the spectroscopic training sample. The marginal redshift and magnitude distributions for the subsets and the combined sample (black) are shown in the histograms (above and right respectively).}
    \label{fig:training_sample}
\end{figure}

In Figure~\ref{fig:training_sample} we present the redshift and F277W magnitude distribution ($m_{\text{F277W}}$) of the combined spectroscopic training sample, split by their origin.
As expected, we see that the CANDELS spec-$z$ compilation predominantly spans brighter magnitudes at lower redshift ($z_{\text{spec}} < 2$).
The \emph{JWST}/NIRSpec sources, although spanning the full redshift range, become the dominant population above $z\sim1.5$ while also spanning almost the full dynamic range in magnitude.
While the additional \emph{JWST}/WFSS sources do not contribute a large fraction of the overall training sample, they provide valuable additional faint $z > 4$ sources that complement the more numerous NIRSpec sources due to being un-targeted and potentially probing novel colour-space. 

\section{Photo-$z$ methodologies}
\label{sec:method}
In the following section we present the specific photo-$z$ methodologies applied in this analysis.
Here and throughout the subsequent analysis, to quantitatively compare the photo-$z$ quality from different methods or training runs we use a number of common literature metrics.
These include a measure of the scatter: 
\begin{equation}
\sigma_{\textup{NMAD}} =1.48 \times \text{median} ( \left | \delta z \right | / (1+z_{\textup{spec}})),
\end{equation}
where $\delta z = z_{\textup{phot}} - z_{\textup{spec}}$ \citep[e.g. ][]{Dahlen:2013eu}.
Similarly, we define the outlier fraction as
\begin{equation}
\text{OLF}_{0.15} = \left | \delta z \right | / (1+z_{\text{spec}}) > 0.15,
\end{equation}
and a measure of the overall sample bias:
\begin{equation}
    \Delta_{z} = \text{median} ( \delta z / (1+z_{\text{spec}}) ).
\end{equation}

\subsection{EAzY Template fits}
To provide a fiducial reference photo-$z$ sample with which to compare our ML-derived estimates, we calculate template-based photo-$z$ estimates in line with what is effectively the standard approach in the field.
In particular, we follow an approach similar to that employed by \citet{williams2024} for the realistic photo-$z$ analysis of simulated galaxies in the PANORAMIC survey paper; we measure photo-$z$s using the EAzY \citep{Brammer2008} template fitting code using the \texttt{sfhz} set supplemented with the obscured AGN template of \citet{Killi2023}.

Since the aperture corrected fluxes used for this analysis do not include full PSF-homogenisation, significant zeropoint corrections are required to produce reliable template-based photo-$z$ estimates. 
EAzY can automatically derive iterative zeropoint corrections using a spectroscopic training sample, but this analysis must include a more complete wavelength coverage than the NIRCam filters used in this analysis in order to accurately constrain the full SED shape and hence calibrate the respective average offsets that must be applied.
We therefore apply fixed zeropoint offsets derived for the same DJA aperture photometry using the GOODS South field for which a full suite of filters is available.\footnote{\url{https://dawn-cph.github.io/dja/blog/2023/07/14/photometric-catalog-demo/}}
Specifically, we apply multiplicative zeropoint corrections of F115W: 0.876, F150W: 0.871, F200W: 0.903, F277W: 1.0, F356W: 1.077, and F444W: 1.148.

Finally, following standard practice within the field, when running EAzY we include an additional 5\% flux uncertainty added in quadrature to the individual flux uncertainties.

\subsection{Gaussian Process - \textsc{GPz}}
The first ML algorithm we use to derive photo-$z$ estimates is the Gaussian process (GP) photometric redshift code, \textsc{GPz} \citep{2016MNRAS.455.2387A}.
\textsc{GPz} provides several benefits over standard GP implementations that are particularly suited to photo-$z$ estimation, including the use sparse of GPs to reduce computational costs with little to no cost to the accuracy of resulting model predictions.
Furthermore, \textsc{GPz} is able to incorporate variable, non-uniform (i.e. heteroscedastic)  noise in the input data, with the train models accounting for both the intrinsic noise within the data \emph{and} model uncertainties due to limitations in the training data.
Finally, \textsc{GPz} also allows for cost-sensitive learning (CSL), allowing different parts of parameter space to be weighted more or less importantly based on the specific scientific requirements.
As in \citet{duncan2022}, we employ the C++ implementation of \textsc{GPz}, \texttt{gpz++}\footnote{\url{https://github.com/cschreib/gpzpp}}, which provides substantial speed improvements and can accommodate missing data as described in \citet{almosallam-thesis}.
In a change from previous applications, in this analysis we run \texttt{gpz++} through a dedicated Python wrapper, \texttt{gpz\_pype}\footnote{\url{https://dunkenj.github.io/gpz_pype/}} that improves convenience and automation.

\citet{2020MNRAS.498.5498H} and \citet{duncan2022} explore a number of augmentations to the \textsc{GPz} photo-$z$ approach, including the use of CSL weighting and the division of the full/training populations into different subsets of parameter space for training and prediction.
While both options can yield improvements in photo-$z$ accuracy in large spectroscopic training samples, the computational benefits for sample division are less pertinent to the small \emph{JWST} training sample employed here.
For simplicity, in this analysis we therefore employ \textsc{GPz} in its most simple practical approach, with training and prediction applied to the full parameter space in one single \textsc{GPz} model with all training sources given equal weight.
For training, we divide the input spectroscopic sample into a 70:20:10 split between training, validation and test samples.

\begin{figure}[t!]
    \centering
    \vspace{0.2cm}
    \includegraphics[width=\columnwidth]{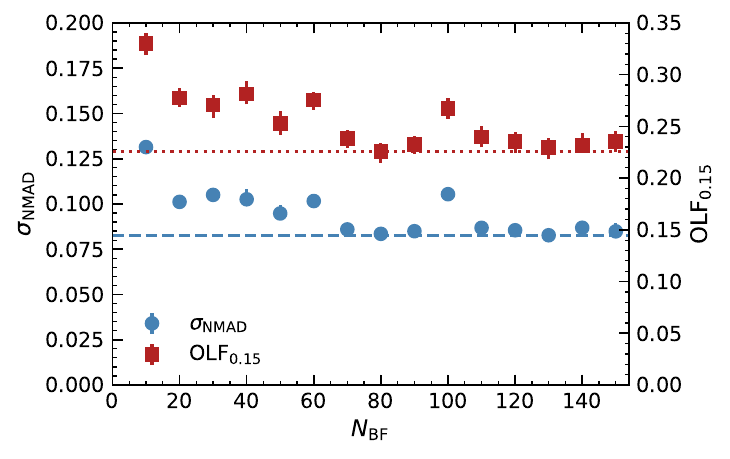}
    \caption{Photo-$z$ scatter ($\sigma_{\text{NMAD}}$) and outlier fraction ($\text{OLF}_{0.15}$) from \textsc{GPz} as a function of basis functions, $N_{\text{BF}}$, used for training. Datapoints and corresponding uncertainties present the median and 16 to 84th percentiles of the $\sigma_{\text{NMAD}}$/$\text{OLF}_{0.15}$ calculated for 100 bootstrap resamples of the test sample. Above $\sim70$ basis functions, the resulting model complexity yields no significant gain in photo-$z$ precision and reliability.}
    \label{fig:nbf}
    \vspace{0.2cm}
\end{figure}

To determine the optimal number of GP basis functions for the data in this analysis, we iteratively train \textsc{GPz} with a range of basis functions in increasing complexity from 10 to 150.
For this experiment, \textsc{GPz} was trained with the covariance between basis functions set to the variable diagonal \citep[`GPVD';][]{almosallam-thesis}.
Figure~\ref{fig:nbf} presents the resulting photo-$z$ scatter ($\sigma_{\text{NMAD}}$) and outlier fraction ($\text{OLF}_{0.15}$) as a function of basis functions used for training.
We find that for $N_{\text{BF}}$ greater than $\sim70-80$, there is no statistically significant improvement in either metric.
We also find that increasing the complexity of the fitting to full variable covariance between basis functions (`GPVC') yields no significant improvement in photo-$z$ outputs for significantly slower training times. 
To avoid over-fitting and minimise the training time requirements, we therefore employ $N_{\text{BF}} = 75$ with variable diagonal covariance matrix for all subsequent \textsc{GPz} training runs. 
We note that should further training samples be added, or additional filters (or training features included), additional basis functions, or more complex covariances could be beneficial and this analysis would need to be reperformed.
Furthermore, the additional division and CSL weight augmentations outlined \citet{2020MNRAS.498.5498H} and \citet{duncan2022} could also yield improvements in other applications.
For this analysis, however, we proceed with the simplest effective application.


\subsection{Nearest neighbour - \textsc{NNpz}}
Finally, while \textsc{GPz} has been successfully applied to a range of datasets \citep[e.g.][]{2020MNRAS.498.5498H,Duncan:21}, it may not be the optimal algorithm for photo-$z$ prediction for datasets with the combination of features and training density explored here.
We therefore also apply a secondary ML photo-$z$ algorithm based on the Nearest-Neighbour Photometric Redshift (NNPZ) algorithm employed by the \emph{Euclid} survey \citep{2020A&A...644A..31E}, although the exact implementation as presented below differs from that described in \citet[][hence we refer to our method as \textsc{NNpz} to differentiate]{2020A&A...644A..31E}.

The first stage of the algorithm finds initial neighbours in flux space using a $k$-dimensional tree based on the Euclidean distance between the input and training features. 
Due to the smaller training sample, we use a smaller initial sample of the 300 nearest neighbours \citep[cf. 2000 in ][]{2020A&A...644A..31E}.
The final neighbours are then found from this initial set using a $\chi^{2}$ metric, allowing for the inclusion of the flux uncertainties in both the input and training samples.
For consistency with the template-based photo-$z$ estimates, we include an additional 5\% flux uncertainty in quadrature when calculating the $\chi^2$ distance metric.

The original `NNPZ' implementation as outlined in \citet{2020A&A...644A..31E} uses a reference training sample consisting of photo-$z$s estimated using deeper and more extensive (i.e. more filters over wider wavelength coverage) photometry; the output photo-$z$ posteriors are generated through a weighted combination of the individual photo-$z$ posteriors of the 30 nearest neighbours.
With the increasing filter coverage of the best-studied \emph{JWST} fields \citep{naidu2024}, such an approach could potentially be applied to predictions for JWST.
However, in the pure parallel \emph{JWST} fields that are the motivation of this paper, the photometric depths reached are often comparable to or in some cases deeper than the current legacy field data.
For this analysis, we therefore generate output photo-$z$ posteriors using the spectroscopic redshifts themselves. 
For each input galaxy, we generate an output posterior using a Gaussian Kernel Density Estimate (KDE) of the 30 best matching neighbours, with each point weighted proportional to $e^{-\chi^{2}/2}$.

The Gaussian KDE estimation approach may struggle to describe the full details of multi-modal photo-$z$ posteriors.
However, as we will demonstrate below, the resulting posteriors are able to provide accurate descriptions of the photo-$z$ uncertainties sufficient for analysis in this context.

\subsection{Photo-$z$ uncertainties and calculation of point-estimates}\label{sec:method-uncertainty}
With each of the three individual photo-$z$ estimates producing photo-$z$ posteriors of a range of complexity, it is valuable to quantify the accuracy and if necessary recalibrate.
We follow the method outlined in \citet[][and originally proposed in \citealt{2016MNRAS.457.4005W}]{Duncan:2017ul} to quantify the photo-$z$ posterior accuracy by calculating the distribution of threshold credible intervals (CI), $c$, where the spectroscopic redshift intersects the redshift posterior.
In the idealised case where e.g. 10\% of galaxies have the true redshift within the 10\% credible interval, 20\% within their 20\% credible interval, etc., the expected distribution of $c$ values should be constant between 0 and 1 \citep[see also the more commonly used probability integral transform, PIT, metric][]{polsterer2016}.
The cumulative distribution, $\hat{F}(c)$, should therefore follow a straight 1:1 relation, i.e. a $Q-Q$ plot.
Curves which fall below this expected 1:1 relation indicate that there is overconfidence in the photometric redshift errors (the posteriors are too sharp) while curves which lie above the 1:1 trend indicate under-confidence with posteriors that are too broad.
The photo-$z$ posteriors can therefore be broadened or sharpened either uniformly \citep{Dahlen:2013eu} or as a function of input properties \citep[e.g. magnitude;][]{duncan2022} to bring the overall $\hat{F}(c)$ distribution closer to the idealised 1:1 relation.

Next, for plotting and the calculation of photo-$z$ metrics derived from point estimates, we follow the approach presented by \citet[][]{Duncan2019} that is designed to accurately condense potentially multi-modal photo-$z$ posteriors into representative `best' photo-$z$ estimates (where simple maximum a posteriori or median values can misrepresent the $P(z)$).
Specifically, after the photo-$z$ posteriors have been calibrated, for each individual $P(z)$ we first calculate the 80\% highest probability density (HPD) CI by starting at the redshift peak probability and lowering a threshold until 80\% of the integrated probability is included.
We then identify the primary peak (and secondary peak if present) by identifying the points where the $P(z)$ cross this threshold.
For each peak, we then calculate the median redshift within the boundaries of the 80\% HPD CI to produce our point-estimate of the photo-$z$ ($z_{1,\textup{median}}$ or $z_{2,\textup{median}}$). 
Unless otherwise specified, for any subsequent figures or quality metrics we define the `best' photo-$z$ as $z_{\text{phot}} = z_{1,\textup{median}}$.
As a measure of the redshift uncertainty, we use the lower and upper boundaries of the 80\% HPD CI peaks (i.e. where the $P(z)$ crosses the threshold), $z_{1,\textup{min}}$ and $z_{1,\textup{max}}$ respectively.

\section{Results}\label{sec:results}
\subsection{Individual photo-$z$ estimates}\label{sec:results-individual}
To ensure like-for-like comparisons between the photo-$z$ methodologies explored here, we use the same test sample across all three methods.
Since no training or optimisation is applied based on the spectroscopic training sample, we run EAzY template fitting for the full training sample.
However, following the methodology outlined above, when training \textsc{GPz} the full spectroscopic training sample is split into a 70:20:10 ratio for training:validation:test.
Once \textsc{GPz} has been trained, we use the resulting splits to also extract the same test subset for the EAzY template fits.
For \textsc{NNpz}, we use the test and validation subsets (i.e. 90\% of sources) as the nearest-neighbour training sample, with predictions generated for the 10\% of sources reserved for testing. 
In total the test sample consists of 1906 sources spanning between $0 \lesssim z \lesssim 13$ with a median redshift of $z\sim2$.

\begin{figure}
    \centering
    \vspace{0cm}
    \includegraphics[width=\columnwidth]{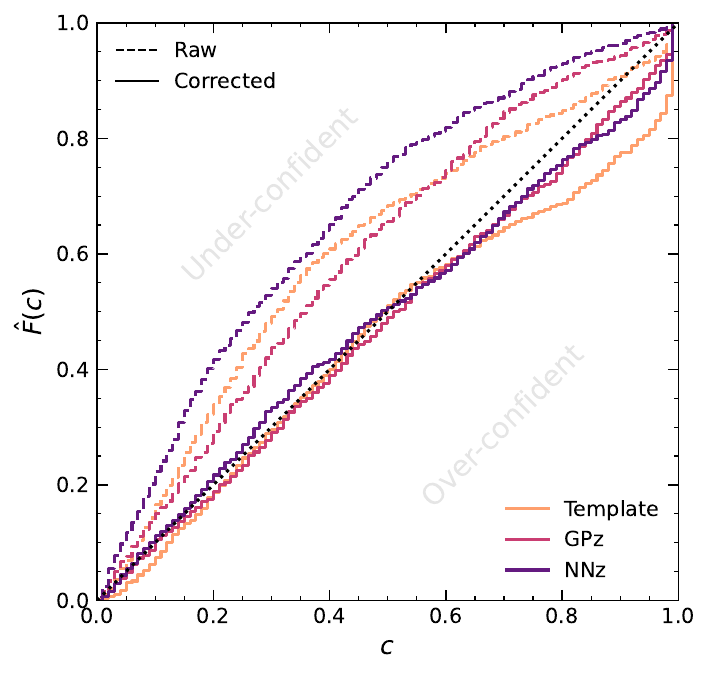}
    \caption{Cumulative distribution of threshold credible intervals, $c$, ($\hat{F}(c)$) for the spectroscopic test sample before (dashed lines) and after (solid lines) uncertainty calibration for each of the individual photo-$z$ methodologies. Lines that rise above the 1:1 relation illustrate under-confidence in the photo-$z$ uncertainties (uncertainties overestimated) while lines that fall under illustrate over-confidence (uncertainties underestimated).}
    \label{fig:posteriors_individual}
\end{figure}

Following prediction for each method, in order to make a fair comparison between the predicted photo-$z$ uncertainties we perform a simple calibration of the posteriors following the methodology outlined in Section~\ref{sec:method-uncertainty}.
For detailed analysis and applications to real science samples, calibration of the posteriors can be done as a quantitative optimisation to minimise the Euclidean distance between the measured and optimal distribution \citep[see e.g.][]{duncan2022} for a retained subset of the test sample.
Given the illustrative nature of the example presented in this analysis, we choose to perform a simple manual adjustment of the overall sample.
In Figure~\ref{fig:posteriors_individual}, we present cumulative distribution of threshold CIs of a subset of the \emph{test} sample for each of the photo-$z$ methods before (dashed lines) and after (solid lines) a basic calibration attempt, where the other 50\% has been used to derive the corrections to be applied.
For the EAzY and \textsc{NNpz} estimates, where the photo-$z$ posteriors are non-Gaussian, we scale the photo-$z$ posteriors such that $P(z)_{\text{Corr}} \propto P(z)_{\text{raw}}^{1/\alpha}$ \citep{Dahlen:2013eu}, where $\alpha < 1$ will result in a sharpening of the posteriors and $\alpha > 1$ results in a broadening.
For our ad hoc corrections, we use $\alpha_{\text{EAzY}} = 0.35$ and $\alpha_{\text{NNpz}} = 0.3$ to sharpen the posteriors.
For GPz, where the posteriors are by definition Gaussian, we simply scale the width of the Gaussian uncertainty by a factor of $2/3$.

\begin{figure*}
    \centering
    \vspace{0cm}
    \includegraphics[width=\textwidth]{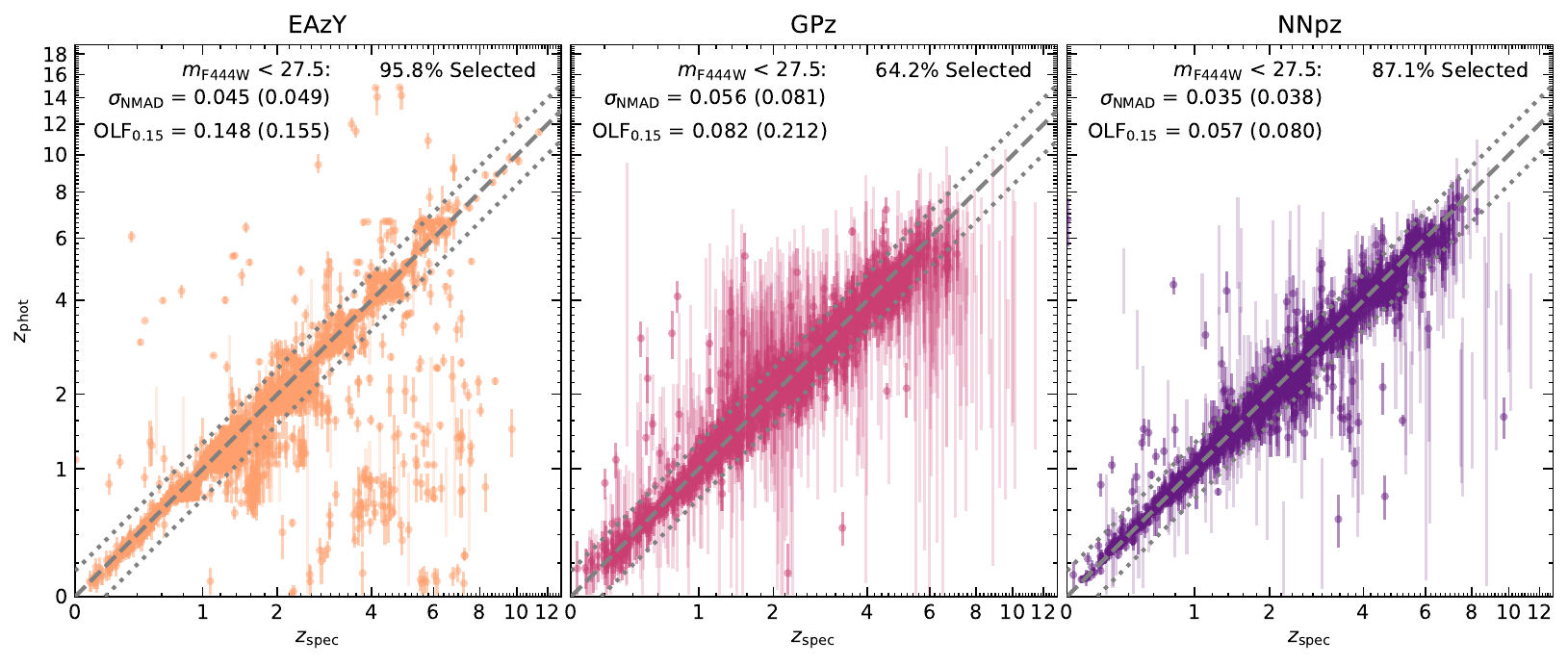}
    \caption{Panels illustrate the best photo-$z$ point-source estimate and corresponding uncertainties (see text for definitions) of each individual methodology for the same spectroscopic test sample. Sources selected as `good', with well constrained primary photo-$z$ peaks, are shown as filled symbols and corresponding error bars with the percentage of test sources selected by this criteria annotated in the upper right corner of each panel. For sources that don't meet these criteria, we plot only the corresponding uncertainty range. For each method, we also show the photo-$z$ scatter and outlier statistics achieved for the `good' photo-$z$ (and full) samples brighter than $m_{\text{F444W}} < 27.5$.}
    \label{fig:outputs_individual}
    \vspace{0.2cm}
\end{figure*}

In Figure~\ref{fig:outputs_individual}, we present the resulting distribution of photo-$z$ versus spec-$z$ for each of the three individual photo-$z$ methods.
To mimic the approach that could be taken for selecting real photo-$z$ samples, we make a distinction between the full set of photo-$z$ predictions and those which would be deemed `good' based solely on the confidence of the photo-$z$.
For this example, we define `good' as sources with a primary photo-$z$ peak that is $z_{\text{max}} - z_{\text{min}} < 0.3\times(1+z_{\text{phot}})$ ($\sigma\approx0.11\times(1+z)$ if the primary peak is a perfect Gaussian).
In Figure~\ref{fig:outputs_individual}, we therefore plot confident photo-$z$ predictions as filled symbols and corresponding uncertainties, while for apparently insecure predictions we plot only the uncertainty range in the background.
The percentage of test sources that meet the selection criteria is shown for each method.
We also calculate photo-$z$ quality statistics for both the `good' and all photo-$z$s for brighter than $m_{\text{F444W}} < 27.5$ \citep[cf.][]{williams2024}.

Exploring the results of the individual photo-$z$ predictions for the 6-band PANORAMIC filter set, we see a number of key trends.
\begin{itemize}
    \item EAzY: In agreement with the typical photo-$z$ performance presented by \citet{williams2024}, the overall quality of the EAzY photo-$z$s are statistically very good for the majority of sources. 
    Across most redshifts probed, the photo-$z$s are accurate and EAzY provides supposedly tight constraints on the primary peak such that 96\% of predictions are deemed to be `good'. 
    However, we observe the some of the same issues identified by \citet{williams2024}, with a number of catastrophic outliers for which EAzY is extremely confident.
    Some of these outliers occur at specific redshift ranges (i.e. $1.5 < z_{\text{spec}} < 2$), while others are distributed more sparsely at $z\gtrsim2$.
    
    Most worringly for the key scientific goals of deep \emph{JWST} parallel surveys, EAzY incorrectly predicts confident photo-$z$s at $z > 9$.
    While the true $z_{\text{spec}}> 9$ sources have accurate and precise predictions, these are substantially outnumbered by the spurious $z_{\text{phot}}>9$ sources.
    Conversely, a significant fraction of sources with true redshifts in the range $4 < z < 8$ are confidently predicted to be at low redshifts ($z < 3$).
    Therefore, while an overall outlier fraction of $\lesssim15\%$ may potentially be acceptable for some science cases, the catastrophic failure of EAzY photo-$z$s in regions of parameter space most relevant to deep \emph{JWST} studies ($z \gtrsim 4$) could significantly limit the full potential of these datasets, either through wasted resources on spectroscopic follow-up of spurious $z > 9$ sources or through incomplete samples of rare populations at $4 < z < 8$.

    We note that EAzY's confidence for a significant fraction of catastrophic failures may seem at odds with the results of Figure~\ref{fig:posteriors_individual} that quantitatively suggests the calibrated uncertainties are accurate.
    However, this discrepancy can be explained by the fact that the EAzY posteriors are highly multi-modal, with  $>50\%$ of predictions having significant secondary photo-$z$ solutions (see also discussion in Section~\ref{sec:results-consensus} and Figure~\ref{fig:pofz_examples}).
    
    \item \textsc{GPz}: The photo-$z$ predictions from \textsc{GPz} are notably different from those produced by EAzY.
    For the full sample of predictions, \textsc{GPz} exhibits significantly higher scatter and outlier fraction ($>20\%$) than EAzY.
    However, the uncertainties on individual predictions are also correspondingly larger, with only $\sim64\%$ of predictions meeting the threshold for confident predictions.
    The large uncertainties for \textsc{GPz} are likely a result of the same colour degeneracies that give rise to multi-modal solutions for EAzY, where the simpler Gaussian posterior has to be much larger to encompass the potential solutions within the 2-3$\sigma$ uncertainties.
    When limiting to only the confident photo-$z$ estimates, \textsc{GPz} yields a photo-$z$ sample with comparable scatter to EAzY but with significantly reduced absolute outlier fraction.
    Crucially, the small but scientifically important catastrophic outlier populations at $z_{\text{phot}}>9$ and $4 < z_{\text{spec}} < 9$ are not present in \textsc{GPz}'s predictions. 
    While \textsc{GPz} is unable to correctly identify the true $z > 8$ populations, the corresponding uncertainties are at least large, preventing selection of spurious $z > 8$ candidates.
    Using \textsc{GPz} photo-$z$ predictions alone, it is therefore not possible to select novel $z > 8$ populations with this limited filter set.
    However, they could be used to select very pure samples of galaxies up to $z\sim7$, with negligible contamination from catastrophic redshift failures.

    \item \textsc{NNpz}: By most metrics, \textsc{NNpz} provides the best photo-$z$ performance for the samples explored here.
    The overall robust scatter and outlier fractions for both the full test sample and the subset for which \textsc{NNpz} is confident are the best of all three methods.
    The fraction of sources selected as `good' for \textsc{NNpz} is $\sim87\%$, lower than for the template-based estimates but significantly higher than \textsc{GPz}.

    Like \textsc{GPz}, \textsc{NNpz} does not identify spurious any $z > 9$ populations, but does struggle to correctly identify all of the true $z > 9$ sources in the test sample.
    \textsc{NNpz}'s predictions also significantly reduce the number of $z_{\text{spec}} < 9$ sources with catastrophic photo-$z$ failures compared to template fitting, including the significant population of failures at $1.5 < z_{\text{spec}} < 2$.
    Crucially, many of the $4 < z_{\text{spec}} < 9$ sources \textsc{NNpz} does fail for have correspondingly large uncertainties and are not selected as `good' estimates.

    Despite being the simplest methodology, the high confidence sample is able to achieve robust scatter $\sigma_{\text{NMAD}}=0.035$ and absolute outlier fraction $\text{OLF}_{0.15}=5.7\%$.
    These metrics demonstrate that \textsc{NNPz} is able to produce photo-$z$ estimates between $0 < z < 7$ with only 6 NIRCam photometry bands that are comparable to those achieved by literature estimates using extensive HST optical to NIR filter sets \citep[e.g.][]{Dahlen:2013eu,Duncan2019b,Kodra2023}, and template estimates for \emph{JWST} survey fields with much more extensive NIRCam (and HST) filter sets \citep[e.g.][]{Adams2024,Duan2025}.
\end{itemize}

Individually, both ML photo-$z$ methods here can potentially offer scientific advantages over the more commonly used template fitting approach, yielding better overall photo-$z$ statistics and the ability to select photo-$z$ samples with significantly higher purity than template-only estimates.
However, none of the methods are optimal, as the improvement in overall statistics for the ML methods at $z \lesssim 8$ comes at the expense of being able to correctly identify the true $z \gtrsim 8$ sources within the sample.
As introduced at the beginning of this manuscript, the ML methods presented here do not necessarily have to provide a direct replacement for template fitting approaches; as multiple works have demonstrated \citep{Dahlen:2013eu, CarrascoKind:2014jg, Duncan:21, Kodra2023}, the statistical combination of multiple photo-$z$ estimates has the potential produce consensus estimates that significantly improve on the results of any individual method.

\subsection{Combined consensus or `hybrid' photo-$z$ predictions}\label{sec:results-consensus}
To produce consensus photo-$z$ estimates, we use Hierarchical Bayesian combination of the individual photo-$z$ posteriors \citep[][see also \citeauthor{Duncan:2017wu}~\citeyear{Duncan:2017wu} and \citeauthor{Duncan2019b}~\citeyear{Duncan2019b} for applications combining EAzY and \textsc{GPz}]{Dahlen:2013eu,CarrascoKind:2014jg}.
Full details of the methodology are presented in the references above, however, in short we define for each individual redshift estimate, $i$,
\begin{equation}
	\begin{split}
	P(z, f_{\text{bad}})_{i} = P(z|\text{bad measurement})_{i}f_{\text{bad}} \\
	+ P(z|\text{good measurement})_{i} (1-f_{\text{bad}}),
	\end{split}
\end{equation}
where $P(z|\text{bad measurement})$ is the redshift probability distribution assumed in the case where the estimated $P_{m}(z)_{i}$ is incorrect and $P(z|\text{good measurement}) \equiv P_{m}(z)_{i}$ is the case where it is correct.
For this analysis, we assume a flat redshift prior for $P(z|\text{bad measurement})$, but note that alternative assumptions such as a comoving volume dependent prior yields comparable results.
The combined $P(z, f_{\text{bad}})$ from all $n$ measurements is then given by
\begin{equation}
	P(z, f_{\text{bad}}) = \prod_{i=1}^{n}P(z, f_{\text{bad}})_{i}^{1/\beta},
\end{equation}
where the hyper-parameter, $\beta$, defines the degree of covariance between the different measurements.
By marginalising over $f_{\text{bad}}$ we can then produce a consensus redshift probability distribution for each object
\begin{equation}
	P(z) = \int^{f^{\text{max}}_{\text{bad}}}_{f^{\text{min}}_{\text{bad}}} P(z, f_{\text{bad}}) df_{\text{bad}},
\end{equation}
where $f^{\text{min}}_{\text{bad}}$ and $f^{\text{max}}_{\text{bad}}$ are additional hyper-parameters defining the lower and upper limits on the fraction of bad measurements.
By construction these are limited to the range $0 \leq f_{\text{bad}} \leq 1$, but the exact limits used when marginalising over $f_{\text{bad}}$ can also be explicitly tuned or informed by the training sample.
For this analysis, we know from the analysis above that the expected fraction of bad measurements (i.e. catastrophic outliers) is of order $\sim10\%$.
We therefore marginalise over the range $0 \leq f_{\text{bad}} \leq 0.1$.

The remaining hyper-parameter that is explicitly tuned is the covariance, $\beta$.
For completely independent estimates $\beta = 1$, while for estimates that are fully covariant $\beta = n$ and is therefore limited to $n = 2$ and 3 in this work depending on which permutation of inputs are used.

In this section we explore three permutations of consensus photo-$z$s; combining the template fitting estimates with each of \textsc{GPz} and \textsc{NNpz} as well as the combination of all three estimates presented here.
As in Section~\ref{sec:results-individual}, given the illustrative nature of the results presented in this analysis we do not perform a detailed optimisation of $\beta$ and instead perform a simple manual adjustment to produce output posteriors for the test sample that closely match the ideal 1:1 distribution of threshold CI.
\begin{figure}[t!]
    \centering
    \vspace{0.5cm}
    \includegraphics[width=0.95\columnwidth]{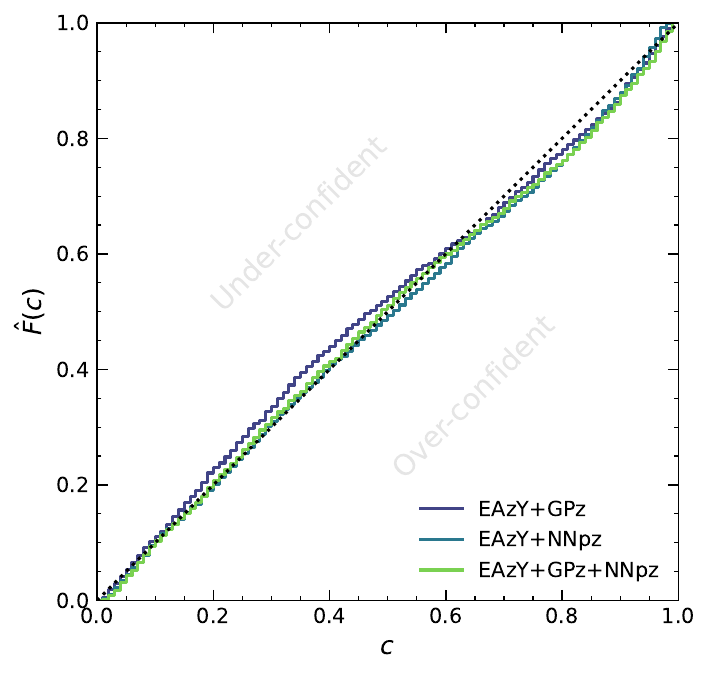}
    \caption{Cumulative distribution of threshold credible intervals, $c$, ($\hat{F}(c)$) for the spectroscopic test sample for eachof the three consensus photo-$z$ estimates derived from Hierarchical Bayesian combination of the template and ML methodologies. Lines that rise above the 1:1 relation illustrate under-confidence in the photo-$z$ uncertainties (uncertainties overestimated) while lines that fall under illustrate over-confidence (uncertainties underestimated).}
    \label{fig:posteriors_consensus}
\end{figure}
\begin{figure*}[t!]
    \centering
    \vspace{0cm}
    \includegraphics[width=\textwidth]{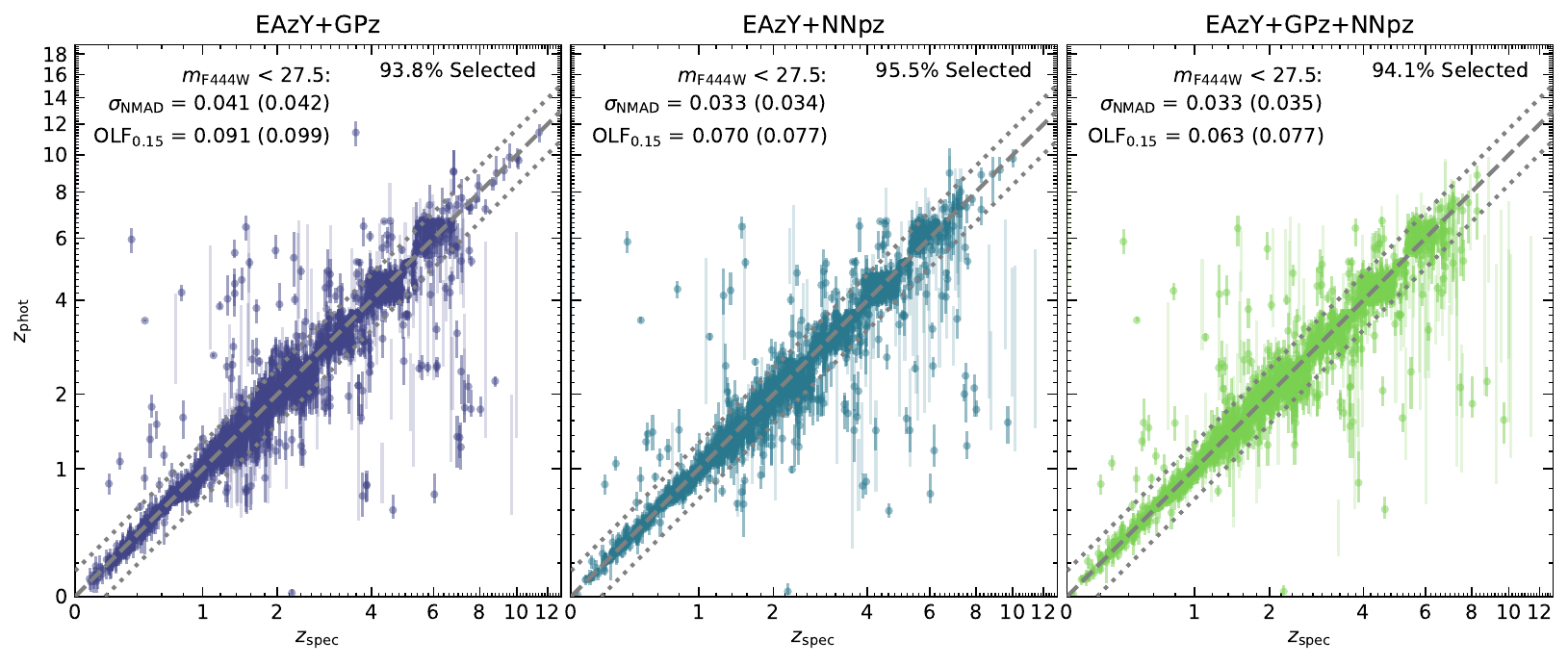}
    \caption{Panels illustrate the best photo-$z$ point-source estimate and corresponding uncertainties (see text for definitions) of the three consensus photo-$z$ estimates derived from Hierarchical Bayesian combination of the template and ML methodologies. Symbols and annotations are as described for Figure~\ref{fig:outputs_individual}.}
    \label{fig:outputs_consensus}
\end{figure*}
The chosen $\beta$ parameters for the three consensus estimates are: EAzY+\textsc{GPz} = 1.5, EAzY+\textsc{NNpz} = 1.7, and EAzY+\textsc{GPz}+\textsc{NNpz} = 2.1.
The resulting distribution of threshold CIs for each estimate is presented in Figure~\ref{fig:posteriors_consensus}.
All three consensus photo-$z$s yield uncertainties that are more accurate than any individual estimate (cf. Figure~\ref{fig:posteriors_individual}), with $\hat{F}(c)$ lying close to the idealised distribution and some visible improvements in the accuracy of the wings of the posteriors (the over-confidence at $c>0.8$).

In Figure~\ref{fig:outputs_consensus} we present the resulting distribution of `best' photo-$z$ estimate for each of the three consensus estimates as well as the corresponding scatter and absolute outlier fraction metrics. 
As for the individual estimates explore above, for each consensus estimate we define a sample of confident or `good' photo-$z$s based on the width of the primary photo-$z$ solution ($z_{\text{max}} - z_{\text{min}} < 0.3\times(1+z_{\text{phot}})$).

All permutations of the consensus estimate are able to produce photo-$z$ estimates that have significantly reduced scatter and outlier fractions compared the EAzY only estimates, while also retaining the high fraction of sources with confident photo-$z$ solutions.
Statistically, the lowest outlier fraction across all estimates is produced by the \textsc{NNpz}-only run.
However, the combination of EAzY+\textsc{NNpz}, or EAzY+\textsc{GPz}+\textsc{NNpz} results in only marginally poorer outlier fraction for `good' photo-$z$ estimates (6.3-7.0\% vs 5.7\%) while also producing confident photo-$z$ solutions for a significantly higher percentage of the population (94-95\% vs 87\%) and therefore a reduced scatter and outlier fraction for the full population when no quality cuts are applied. 

\begin{figure}
    \centering
    \vspace{0cm}
    \includegraphics[width=\columnwidth]{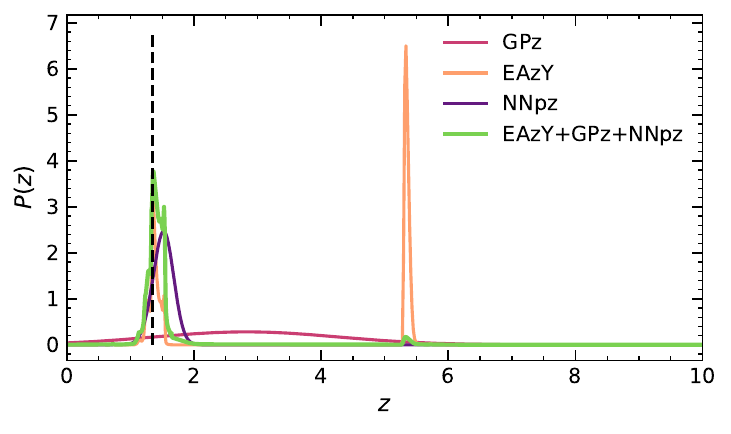}
    \includegraphics[width=\columnwidth]{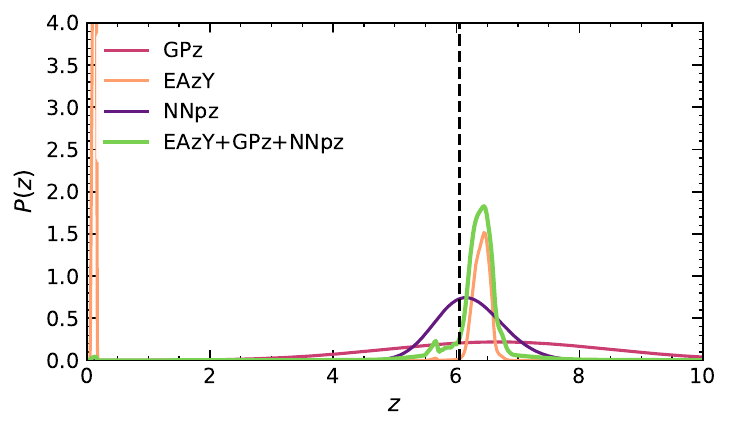}
    \caption{Photo-$z$ posteriors for individual estimates and the combined consensus estimate for examples where incorporating ML estimates provide more accurate and precise results than any individual method. Vertical dashed lines illustrate the correct $z_{\text{spec}}$ for each source.}
    \label{fig:pofz_examples}
\end{figure}

To illustrate how the consensus estimates are able to achieve significant improvements in these cases, in Figure~\ref{fig:pofz_examples} we present the respective individual photo-$z$ posteriors for two representative example.
In both examples, we see that EAzY produces highly multi-modal solutions due to the large number of potential colour degeneracies from the limited filter set.
However, in both cases at least one of the potential solutions lies close to the true redshift that the ML estimates identify correctly (albeit with potentially lower precision).

In Figure~\ref{fig:outputs_consensus} we can see that for some of the catastrophic failures in template fitting, the additional information from the more accurate but less precise ML-based photo-$z$ is not sufficient to push the consensus estimate to the correct solution.
However, for many of the most severe template fitting failures, the consensus estimates are significantly improved.
For both EAzY+\textsc{GPz} and EAzY+\textsc{NNpz}, the consensus estimates are also able to accurately predict the redshifts for a significant fraction of the true $z > 8$ populations.
For the EAzY+\textsc{GPz}+\textsc{NNpz} estimates, the best photo-$z$ estimates for the $z_{\text{spec}} >8$ sources do drop to below $z\lesssim6$, suggesting that the relative weight of the two ML estimates (that underestimate the redshift) combined becomes sufficient to outweigh the correct prediction from template fitting.

\subsection{Redshift dependence of photo-$z$ quality statistics}
In the section above, we have only examined photo-$z$ metrics derived over the full test sample alongside a qualitative assessment of the photo-$z$ performance at different redshifts.
While the sensitivity of \emph{JWST} now enables training sources over almost the full observable redshift and magnitude range (see Figure~\ref{fig:training_sample}), the overall statistical weight of the sample is strongly affected by overall statistical gains at $z < 2$.
It is therefore informative to compare explore the photo-$z$ metrics as a function of $z_{\text{spec}}$ to quantitatively assess how the ML compare with EAzY and verify that the quantitative improvements seen in the consensus estimates is not a result of improvements in only limited parts of parameter space.

\begin{figure}[t!]
    \centering
    \vspace{0cm}
    \includegraphics[width=\columnwidth]{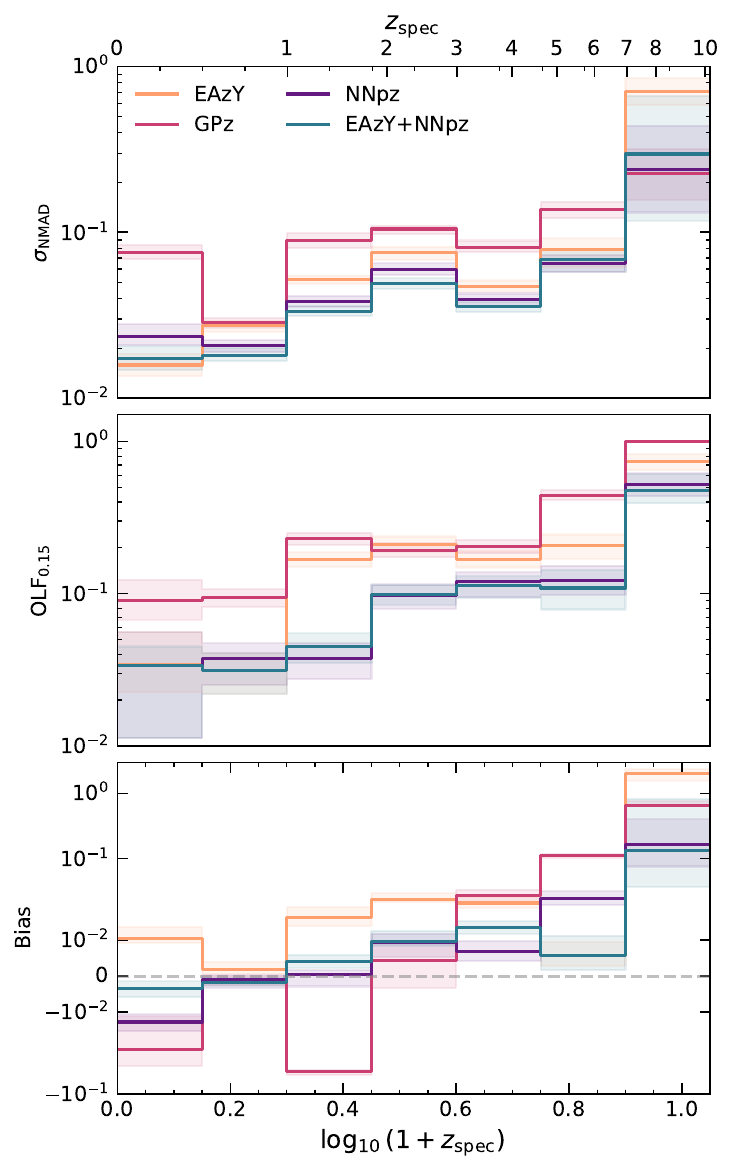}
    \caption{Measured photo-$z$ scatter ($\sigma_{\text{NMAD}}$; top), outlier fraction ($\text{OLF}_{0.15}$; middle) and bias (bottom) as a function of spectroscopic redshift for the three individual photo-$z$ methods as well as one of the best performing hybrid consensus approaches (EAzY+\textsc{NNpz}). Solid lines show the median value for 100 bootstrap resamples, with shaded regions showing the corresponding 16-84th percentile range.}
    \label{fig:stats_vs_z}
\end{figure}

In Figure~\ref{fig:stats_vs_z} we present the redshift dependence of the standard scatter and outlier statistics explored above, along with the addition of a measure of sample bias.
To simplify the comparison, we present statistics for the three individual estimates as well as the best performing consensus estimate when no quality selection criteria are applied (EAzY+\textsc{NNpz}).
To ensure like-for-like comparison in each redshift bin we calculate the photo-$z$ metrics for all sources that are $m_{\text{F444W}} < 27.5$ within a given redshift bin, with no cuts on photo-$z$ quality applied; the statistics are therefore derived from the same subset of galaxies for each method.
Additionally, to help quantify the impacts of the potentially small samples within each bin, we calculate the statistics for 100 bootstrap resamples of each bin; solid lines then indicate the median across all samples with the shaded regions illustrating the corresponding 16 to 84 percentiles.

We find that the consensus estimates yield photo-$z$s that have the lowest scatter/outlier fractions at all redshifts within the uncertainties from sample variance, while also yielding the lowest absolute bias in most redshifts.
Overall, EAzY+\textsc{NNpz}, is able to not only match the best of the individual estimates that went into it, but instead improving on them.

\section{Discussion}\label{sec:discussion}
For the specific combination of data and training samples used in this analysis, the best overall photo-$z$ performance is achieved by either \textsc{NNpz} alone or the combination of EAzY+\textsc{NNpz}/EAzY+\textsc{GPz}+\textsc{NNpz} depending on the particular metric explored or the redshift range of interest.
Nevertheless, the significant improvement over template fitting alone offered by the inclusion of predictions from only \textsc{GPz} demonstrates that even relatively imprecise ML photo-$z$ predictions can yield valuable improvements and that the use of hybrid photo-$z$ estimates for \emph{JWST} is agnostic to the specific ML algorithms employed.
Furthermore, tests with alternative reduced \emph{JWST} filter combinations such as those in the North ecliptic pole EXtragalactic Unified Survey \citep[NEXUS;][F090W, F115W, F150W, F200W, F356W, F444W]{NEXUS2024} yield qualitatively identical results but with small variations in statistical metrics, with \textsc{NNpz} alone able to match or improve on EAzY template fits and further overall improvements to be gained from combining multiple estimates.
The same is true for alternative choices of photometry, for example analysis using model fitting photometry for a reduced subset of the DJA imaging \citep{genin2025} again yields qualitatively similar results.

We note that for both aperture and model-fitting photometry, the inclusion of size features (e.g. flux radius or S\'{e}rsic radius and index) in the ML training yields no significant improvement results for those methods when compared to the improvements in key areas offered by combining multiple estimates. 
However, as training samples continue to grow and more complex morphological features are routinely catalogued for \emph{JWST} samples \citep{kartaltepe2023, ganapathy2025}, the inclusion of additional morphological information could still yield further improvements in ML estimates for these sources \citep[see e.g.][for applications at lower redshift]{2018MNRAS.475..331G, duncan2022}.

\subsection{Strengths and weaknesses of ML and hybrid approaches}
The overall benefits of using consensus estimates has been observed previously in deep galaxy survey samples, either from template approaches alone \citep{Dahlen:2013eu, Kodra2023} or from hybrid ML plus template fitting \citep[e.g.][]{Duncan2019b,Duncan:21}.
The consistent statistical and qualitative improvements shown here, however, unambiguously demonstrate the enormous advantage now offered by ML or hybrid approaches for deep \emph{JWST} imaging surveys with limited filter sets.

One common advantage for template fitting photo-$z$s in deep pencil beam surveys, where filter coverage across the field can vary significantly, is that it can account for missing or variable filters automatically by simply only including available filters in the calculation of $\chi^{2}$ statistics.
In contrast, ML methods must be trained to explicitly predict missing inputs \citep[e.g.][for \textsc{GPz}]{almosallam-thesis} or multiple models must be trained to account for the necessary permutations, thereby potentially adding significantly to computational costs for training and predictions.
However, for the dataset and methodologies presented here, the computational requirements for the additional ML training and prediction steps are effectively negligible for the small samples (of order seconds on a 10-core Apple M1 Pro chip).
Training multiple \textsc{GPz} and \textsc{NNpz} models for all of the filter combinations in the PANORAMIC \citep{williams2024} and BEACON \citep{morishita2025} pure parallel surveys for example is therefore tractable and not a significant disadvantage to this approach.
Alternatively, as demonstrated by the improvement offered by EAzY+\textsc{GPz}, using a smaller but more homogeneous filter set for \textsc{NNpz} at the expense of maximum ML performance would still likely yield significant improvements.
In this regard, we argue that \textsc{NNpz} being fast, simple and easily interpretable offers significant advantages over other proposed hybrid approaches \citep{2017MNRAS.466.2039C, tanigawa2024}.

A caveat that we must acknowledge is that while the training sample now available spans almost all possible redshifts and continuum magnitudes, it is however still incomplete.
Predictions for galaxies with colours outside the available parameter space are therefore poor.
A strength of template fitting is that it is able to leverage our astrophysical knowledge about stellar populations and processes in the inter-galactic medium \citep[e.g. the Gunn-Peterson trough;][]{Madau:1999kl} to make predictions for galaxies beyond the current training samples, as illustrated by the successful confirmation of new high-$z$ samples \citep{curtis-lake2023,arrabalharo2023,Carniani2024}.
Provided the ML estimates provide accurate uncertainties for these cases, it is this complementarity between the two methods however that is a key potential strength of the hybrid approach.

We have demonstrated in Section~\ref{sec:results} that despite any remaining limitations in training samples, ML approaches are already competitive with template fitting.
As training samples continue to grow through large NIRSpec campaigns \citep{Capers2024,Shen2024} and wide-area slitless spectroscopy \citep[e.g.][]{cosmos3d}, these methods are only going to improve.
Similarly, with the filter coverage of the best studied \emph{JWST} fields increasing thanks to the addition of medium \citep{suess2024, muzzin2025} and narrow-band \citep{duncan2025} filters, the photometric precision for sources in these fields is reaching near spectroscopic levels for highly complete samples \citep[see e.g.][]{naidu2024}.
Although not explored here, methods such as \textsc{NNpz} could also still be applied in the same way as presented by \citep{2020A&A...644A..31E}, whereby the output posteriors are constructed from nearest-neighbour matching to template photo-$z$s derived from the full 15-20 band photometry. 

\section{Summary and conclusions}
\label{sec:concl}
We have presented a comprehensive study on the application of machine learning (ML) techniques to improve photometric redshift estimates for galaxies observed in deep JWST surveys with limited filter sets. 
Specifically we have investigated two ML-based methods using Gaussian Processes (\textsc{GPz}) and Nearest Neighbour (\textsc{NNpz}) based approaches, comparing their performance to traditional template-based estimates using the EAzY photo-$z$ code \citep{Brammer:2008gn}.
We also investigate the improvements that can be gained from the combination of multiple methods through Hierarchical Bayesian combination to produce hybrid consensus photo-$z$ estimates.

Our key conclusions can be summarized as follows:
\begin{itemize}
    \item When compared to template fitting, ML based methods can reduce the scatter and outlier fraction for photo-$z$s derived from 6 filters representative of current pure parallel and wide-area \emph{JWST}/NIRCam surveys, with \textsc{NNpz} method outperforming both EAzY and \textsc{GPz}.
    \item The ML methods are particularly effective in reducing catastrophic failures, where $z_{\text{phot}}$ is significantly over- ($z_{\text{phot}} \gtrsim z_{\text{spec}} + 3$) or under-predicted ($z_{\text{phot}} \lesssim z_{\text{spec}} - 2$), scenarios that are particularly relevant for deep \emph{JWST} surveys 
designed to yield complete statistical samples of galaxies in  $z > 2$.
    \item Combining the ML estimates with template-based estimates can lead to even better performance in multiple metrics, with the combination of EAzY+\textsc{GPz}+\textsc{NNpz} producing consensus estimates for sources brighter than $m_{\text{F444W}} < 27.5$ with robust scatter of $\sigma_{\text{NMAD}}=0.033$ and an outlier fraction of $\text{OLF}_{0.15}=0.063$.
\end{itemize}

Overall, our results demonstrate the potential of simple and easily applicable ML-based photo-$z$ techniques to enhance the scientific return of \emph{JWST} datasets with limited filter sets.
These results are particularly relevant for large pure parallel surveys such as PANORAMIC \citep{williams2024} and BEACON \citep{morishita2025} that now constitute a significant fraction of \emph{JWST}'s total survey area and hence offer enormous legacy science value.
For example, through enabling the discovery of extreme sources that may be too rare for standard legacy survey fields \citep[e.g.][]{Xiao2025}, or for providing large complete samples that reduce the impacts of cosmic variance on constraints of key population statistics \citep[as achieved for similar HST pure parallel surveys, e.g.][]{morishita2018, roberts-borsani2022}.
Maximising the potential scientific discovery from these existing NIRCam datasets is also crucial for ensuring the future support of \emph{JWST} offering pure parallel survey modes.

To ensure that the results and methods explored in this analysis can be easily applied to new datasets, or extended through the addition of improved training samples, all analysis steps needed to reproduce the results and associated figures in this manuscript are made available through a persistent DOI at \url{https://doi.org/10.5281/zenodo.17535452}.

\section*{Acknowledgments}
KJD acknowledges support from the Science and Technology Facilities Council (STFC) through an Ernest Rutherford Fellowship (grant number ST/W003120/1).
Some of the data products presented herein were retrieved from the Dawn \emph{JWST} Archive (DJA). DJA is an initiative of the Cosmic Dawn Center (DAWN), which is funded by the Danish National Research Foundation under grant DNRF140.
We declare that all core analysis, code and manuscript text was written by the author, but note that generative AI tools (Github Copilot) were employed to improve and add documentation for the accompanying code repository to help maximise the reproducibility. 

\bibliographystyle{mnras}


\bibliography{mlz_for_jwst}




\end{document}